# EQUILIBRIUM STATE OF A SELF-INTERACTING SCALAR FIELD IN THE DE SITTER BACKGROUND

Alexei A. STAROBINSKY[1,2] and Jun'ichi YOKOYAMA[1]

[1] *Yukawa Institute for Theoretical Physics, Kyoto University, Uji 611 (Japan)*
[2] *Landau Institute for Theoretical Physics, Kosygina St. 2, Moscow 117334 (Russia)*
(July 6, 1994)

## Abstract

Behaviour of a weekly self-interacting scalar field with a small mass in the de Sitter background is investigated using the stochastic approach (including the case of a double-well interaction potential). Existence of the de Sitter invariant equilibrium quantum state of the scalar field in the presence of interaction is shown for any sign of the mass term. The stochastic approach is further developed to produce a method of calculation of an arbitrary anomalously large correlation function of the scalar field in the de Sitter background, and expressions for the two-point correlation function in the equilibrium state, correlation time and spatial physical correlation radius are presented. The latter does not depend on time that implies that the characteristic size of domains with positive and negative values of the scalar field remains the same on average in the equilibrium state in spite of the expansion of the $t = const$ hypersurface of the de Sitter space-time.

# I. INTRODUCTION

De Sitter space-time is one of the most fundamental and symmetric space-times. It is a constant-curvature space-time completely characterized by only one constant $H$, and its group of symmetries is as large as the Poincare group of symmetries of the Minkowski space-time (10 generators). That is why it is so significant to get a clear understanding of behaviour of quantum fields including interacting ones in this background. This program was launched in the pioneer paper by Chernikov and Tagirov [1] for the case of a free massive scalar field (see also [2]- [7] and other papers on this subject). Also, quantum field theory in the de Sitter background has extremely important cosmological applications in connection with the inflationary scenario of the early Universe and generation of scalar perturbations (both adiabatic and isocurvature ones) and gravitational waves from vacuum quantum fluctuations during an inflationary stage.

Due to the high symmetry of the de Sitter background it is natural to define the background (or "vacuum") state of any quantum field as a state vector invariant under action of the full group of symmetries of the de Sitter space-time $O(4,1)$. Note that, due to the possibility of a static representation of a part of the de Sitter space-time surrounded by the event horizon, the same state, if exists, may be also called a "thermal" state with the temperature $T = H/2\pi$ [8]. This fact, however, simply reflects the symmetries of the de Sitter space-time and does not mean that average values of operators in this state, e.g. $\langle\phi^2\rangle$, are given by corresponding thermal values in the flat space-time with the Gibbons-Hawking temperature (in fact, the former are much larger in the case we are interested in). One may also expect that this background state is a stable attractor for an open set of other quantum states in the Schrödinger picture, in other words, that it is a stable equilibrium state. In the Heisenberg picture, this means that average values of all possible operators in an open set of constant state vectors approach those in the equilibrium state as $|t| \to \infty$.

The investigation of a free massive scalar field in the de Sitter background has shown that:
1) such an equilibrium background state does exist if $m^2 > 0$, where $m^2$ is the effective mass squared in the de Sitter background including a contribution from the non-minimal coupling term $\xi R\phi$ in the dynamical equation for the scalar field (if such term is present), $R$ being the scalar Ricci curvature;
2) there is no de Sitter invariant state for $m^2 \leq 0$;
3) in the case $m = 0$, if the de Sitter (inflationary) stage begins (or is "switched on") at the moment $t = 0$, the average value $\langle\phi^2\rangle$ in the Heisenberg state which was the vacuum state for $t < 0$ (according to, say, the quasi-adiabatic definition of vacuum) and in an open set of close states grows as $H^3 t/4\pi^2$ for $Ht \gg 1$ [4–6];
4) for $m^2 > 0$, $m \ll H$, the value of $\langle\phi^2\rangle$ in the de Sitter-invariant state is anomalously large as compared to the corresponding thermal value with the Gibbons-Hawking temperature, namely, $\langle\phi^2\rangle = 3H^4/8\pi^2 m^2$ [3].

Now the question arises what happens in the case of a self-interacting scalar field in the de Sitter background. The most natural type of such an interaction is the quartic coupling $\lambda\phi^4/4$ with $0 < \lambda \ll 1$ in the Lagrangian density of the scalar field. This problem was first considered in [2]. More detailed recent investigations [9,10] showed that there exists a breakdown in the perturbation expansion around the free-field ($\lambda = 0$) operators due to infrared divergences if $m^2 \leq 27H^2/16$. This shows that the field theory has a completely non-trivial infrared structure for these values of $m^2$ (especially in the most interesting case $|m^2| \ll H^2$) that lies beyond any finite order of the perturbation expansion in $\lambda$.

Fortunately, there exists a method to manage this difficulty and to solve the problem in the case $|m^2| \ll H^2$ that uses specific properties of the de Sitter space-time (so it cannot help infrared problems in the flat space-time). This is the stochastic approach to inflation (also called "stochastic inflation") that is based on the idea first proposed in [5] that the infrared part of the scalar field may be considered as a classical (i.e, $c$-number) space-dependent stochastic field which satisfies a local (i.e., not containing spatial derivatives) Langevin-like equation. The final form of the corresponding Fokker-Planck equation for the probability distribution function (PDF) of this classical field which is suitable for our purposes was derived in [11,12] (a partial form of the latter equation for the case of a massless non-interacting scalar field was presented in [13]). In this paper, we make a further step and expand this method to the calculation of two-point and higher-point correlation functions that enables to find any anomalously large Green function of the scalar field in the equilibrium background quantum state (large as compared to the corresponding thermal quantity in the flat space-time). Such functions are generally average values of arbitrary functions of the scalar field itself taken at different points in the de Sitter space-time. On the other hand, any non-zero average values including second time derivatives, squares of first time derivatives and spatial gradients of the field are not anomalously large and cannot be exactly calculated by the method used.

So, the structure of the paper as follows. In Sec. 2, we present basic field equations and make a calculation of the equilibrium value of $\langle \phi^2 \rangle$ using the Hartree-Fock (or Gaussian) approximation to get a crude estimation of this quantity. In Sec. 3, we introduce the stochastic approach, display the Fokker-Planck equation and find a static equilibrium solution for the one-point PDF. Expressions for the temporal autocorrelation function of the scalar field in different cases are derived in Sec. 4. In Sec. 5, the general two-point function is found and its de Sitter invariance in the equilibrium state is proved. Sec. 6 contains discussion of a general picture of fluctuations of the self-interacting scalar field in the de Sitter background.

## II. BASIC EQUATIONS AND THE SCALAR FIELD EVOLUTION IN THE HARTREE-FOCK APPROXIMATION

The Lagrangian density of a massive scalar field with the quartic self-interaction is ($c = \hbar = 1$ and the Landau-Lifshits sign conventions are used throughout):

$$\mathcal{L} = \frac{1}{2}(\phi_{,\mu}\phi^{,\mu} - m_0^2 \phi^2 + \xi R \phi^2) - \frac{\lambda \phi^4}{4} \ . \tag{1}$$

In the de Sitter background, the scalar curvature $R = -12H^2$, $H = const$, so the effective mass in the limit $\lambda = 0$ is $m^2 = m_0^2 + 12\xi H^2$. Thus, the wave equation for the scalar field in this background is

$$\phi^{;\mu}_{;\mu} + m^2 \phi + \lambda \phi^3 = 0 \tag{2}$$

which is also the equation for the corresponding Heisenberg field operator, here ; denotes the covariant derivative. We assume further that $|m^2| \ll H^2$. Properties of the quantum field theory with the Lagrangian density (1) at energies $E \gg H$ essentially coincide with those in the flat space-time because corresponding modes of the field do not feel the space-time curvature. So, this theory is renormalizable but it possesses the Landau "zero-charge" pole at very high energies. Thus, it may exist (both in the flat and curved space-time) as a part of a more fundamental underlying theory only. Also, $\lambda$ is running with energy $E$ for $E \gg H$ according to the flat space-time renormalization-group equation. The de Sitter curvature $H$ produces an effective infrared cut-off for all logarithmic radiation corrections (see, e.g., [5]), so by $\lambda$ we shall further mean $\lambda(H)$. To avoid problems with large radiation corrections we assume that $0 < \lambda(H) \ll 1$ (then $\lambda(0)$ is even less) and that the energy $H$, though possibly being high and even not far from the Planckian one $E_{Pl} = M_{Pl} = 1/\sqrt{G}$, is much lower than the energy of the Landau pole that is exponentially large.

We take the metric of the de Sitter space-time in the form

$$ds^2 = dt^2 - a_0^2 e^{2Ht} d\mathbf{x}^2 = \frac{1}{(H\eta)^2}(d\eta^2 - d\mathbf{x}^2) \quad , a_0 = const \tag{3}$$

where $\eta$ is the conformal time defined by $a_0 e^{Ht} = 1/(-H\eta)$ for $-\infty < \eta < 0$. This metric covers only one half of the whole de Sitter space-time. This is, however, sufficient for our purpose because if a de Sitter invariant state is constructed in a part of the space-time, it may be analytically continued to the whole space-time.

As a first step to the understanding of the behaviour of the interacting scalar field (1) in the de Sitter background, let us calculate the evolution of $\langle \phi^2 \rangle$ in the Hartree-Fock approximation following [4,5] (for a recent discussion of this method, see [14]). This approximation is also called Gaussian because the wave functional in the Schrödinger representation is assumed to be Gaussian. For us, however, it will be more convenient to work in the Heisenberg representation. The Hartree-Fock approximation is known to be exact in the non-interacting case ($\lambda = 0$). Also, it becomes exact if we consider a multiplet of $N$ scalar fields with the $O(N)$ symmetry and take the limit $N \to \infty$ (the so-called $1/N$ expansion). For $\lambda \neq 0$ and $N = 1$, one may expect to get a qualitatively correct result only (if at all). Still the Hartree-Fock approximation is so widely used that it is instructive to see what it produces in our case.

To find the evolution equation for $\langle \phi^2 \rangle$, let us multiply Eq. (2) (considered as the operator equation) by $\phi$ and then take its average in some state vector. We get

$$\frac{1}{2}\langle \phi^2 \rangle^{;\mu}_{;\mu} = \langle \phi_{,\mu} \phi^{,\mu} \rangle - m^2 \langle \phi^2 \rangle - \lambda \langle \phi^4 \rangle \; . \tag{4}$$

It is known from the analysis of the free massless case [4-6] that the main contribution to $\langle \phi^2 \rangle$ is made by long-wave, slowly changing with time modes. So, only the term $3H\frac{\partial}{\partial t}$ may be kept from the d'Alambertian operator in the left side of Eq. (4). The first term in the right side of Eq. (4) contains two derivatives and does not have an anomalously large infrared part. Let us use the Hartree-Fock (or Gaussian) approximation to estimate the last term in Eq. (4). If we assume that $\langle \phi \rangle = 0$ initially, it will be formally zero at all times due to the reflection symmetry $\phi \to -\phi$. Then we get $\langle \phi^4 \rangle = 3\langle \phi^2 \rangle^2$ as if the operator $\phi$ would be a Gaussian stochastic quantity. So, Eq. (4) reduces to

$$\frac{\partial}{\partial t}\langle \phi^2 \rangle = \frac{H^3}{4\pi^2} - \frac{2m^2}{3H}\langle \phi^2 \rangle - \frac{2\lambda}{H}\langle \phi^2 \rangle^2 \tag{5}$$

where the first term in the right side is introduced to recover the correct linear growth of $\langle\phi^2\rangle$ in the limit $\lambda \to 0$. It originates from the time dependence of the upper cut-off for the large infrared part of $\langle\phi^2\rangle$ in the space of conformal wave vectors $k : k \sim aH$, that corresponds to a time-independent cut-off in the physical space. On the other hand, the lower cut-off is constant in the $k$-space because it is determined by the moment when the de Sitter expansion begins (cf. [4,5]). It is clear that all solutions of Eq. (5) approach a constant equilibrium value at $t \to \infty$. In particular, $\langle\phi^2\rangle \to H^2/\pi\sqrt{8\lambda} \approx 0.113\ H^2/\sqrt{\lambda}$ if $m^2 = 0$. This corresponds to the appearance of the effective mass $m^2_{eff} \sim H^2\sqrt{\lambda}$ due to self-interaction.

Certainly, we cannot expect that the Hartree-Fock approximation, being exact for free quantum fields, produces a quantitatively correct result in the presence of self-interaction. In particular, the form of the PDF of $\phi$ that it assumes is completely misleading, it is not Gaussian at all. So, let us turn to the quantitatively correct treatment of the problem now.

## III. STOCHASTIC APPROACH AND THE FOKKER-PLANCK EQUATION

In the stochastic approach [11,12] we represent the Heisenberg operator of the quantum field $\phi$ as

$$\phi(\mathbf{x},t) = \overline{\phi}(\mathbf{x},t) + \int \frac{d^3k}{(2\pi)^{\frac{3}{2}}} \theta(k - \epsilon a(t)H) \left[ a_{\mathbf{k}}\phi_{\mathbf{k}}(t)e^{-i\mathbf{k}\mathbf{x}} + a^\dagger_{\mathbf{k}}\phi^*_{\mathbf{k}}(t)e^{i\mathbf{k}\mathbf{x}} \right]. \tag{6}$$

Here $\overline{\phi}(\mathbf{x},t)$ is a coarse-grained or a long wavelength part of $\phi$, $\theta(z)$ is the step function and $\epsilon$ is a small constant parameter which in fact is not arbitrarily small (we shall return to this point below and determine the inequality it should satisfy). Note that if we consider a quasi-de Sitter background with a slowly changing curvature $|\dot{H}| \ll H^2$, then $\epsilon H$ should be taken as a constant to avoid the appearance of additional (though small) terms in the Langevin and the Fokker-Planck equations written below. So, $\overline{\phi}$ is the field $\phi$ averaged over a constant *physical* $3D$ volume slightly larger than the volume inside the event horizon.

The short wavelength counterpart is characterized by the mode functions $\phi_{\mathbf{k}}(t)$ with $a_{\mathbf{k}}$ and $a^\dagger_{\mathbf{k}}$ being annihilation and creation operators, respectively. It is small, so it satisfies the linear massless equation

$$\ddot{\phi}_{\mathbf{k}} + 3H\dot{\phi}_{\mathbf{k}} + \frac{\mathbf{k}^2}{a^2(t)}\phi_{\mathbf{k}} = 0, \tag{7}$$

which is solved as

$$\phi_{\mathbf{k}} = \sqrt{\frac{\pi}{4}}H(-\eta)^{\frac{3}{2}}H^{(1)}_{\frac{3}{2}}(-k\eta) = \frac{H}{\sqrt{2k}}\left(\eta - \frac{i}{k}\right)e^{-ik\eta}, \tag{8}$$

with $H^{(1)}_{\frac{3}{2}}(z)$ being the Hankel function of the first class. With the above choice of mode functions, the constant Heisenberg quantum state annihilated by $a_{\mathbf{k}}$'s corresponds to the usual adiabatic (or Minkowski) vacuum in the limit $\eta \to -\infty$.

From the equation of motion of the scalar field with some interaction potential $V(\phi)$, $\Box\phi + V'(\phi) = 0$, we obtain the following equation for the slowly-varying coarse-grained part $\overline{\phi}$:

$$\dot{\overline{\phi}}(\mathbf{x},t) = -\frac{1}{3H}V'(\overline{\phi}) + f(\mathbf{x},t), \tag{9}$$

where $f(\mathbf{x},t)$ is given by

$$f(\mathbf{x},t) = \epsilon a(t)H^2 \int \frac{d^3k}{(2\pi)^{\frac{3}{2}}} \delta(k - \epsilon a(t)H) \left[ a_{\mathbf{k}}\phi_{\mathbf{k}}(t)e^{-i\mathbf{k}\mathbf{x}} + a_{\mathbf{k}}^{\dagger}\phi_{\mathbf{k}}^{*}(t)e^{i\mathbf{k}\mathbf{x}} \right]. \tag{10}$$

Eq. (9) may be regarded as the Langevin equation for the stochastic quantity $\overline{\phi}$ (the fact that it is still an operator becomes irrelevant) with a stochastic noise term $f(\mathbf{x},t)$ whose correlation properties are given by

$$\langle f(\mathbf{x}_1,t_1)f(\mathbf{x}_2,t_2)\rangle = \frac{H^3}{4\pi^2}\delta(t_1-t_2)j_0(\epsilon a(t)H|\mathbf{x}_1-\mathbf{x}_2|), \quad j_0(z) = \frac{\sin z}{z}. \tag{11}$$

We note that the white-noise property of the above correlation is preserved even if we take into account the first-order correction arising from self-interaction.

It is straightforward to derive the Fokker-Planck equation for the one-point (or one-domain) PDF $\rho_1[\overline{\phi}(\mathbf{x},t) = \varphi] \equiv \rho_1[\varphi(\mathbf{x},t)] \equiv \rho_1(\varphi,t)$. It has the form:

$$\frac{\partial \rho_1[\varphi(\mathbf{x},t)]}{\partial t} = \frac{1}{3H}\frac{\partial}{\partial \varphi}\left\{V'[\varphi(\mathbf{x},t)]\rho_1[\varphi(\mathbf{x},t)]\right\} + \frac{H^3}{8\pi^2}\frac{\partial^2 \rho_1[\varphi(\mathbf{x},t)]}{\partial \varphi^2} \equiv \Gamma_{\varphi}\rho_1[\varphi(\mathbf{x},t)]. \tag{12}$$

Once we have found a solution of the above equation, we can calculate the expectation value of any function of $\overline{\phi}(\mathbf{x},t)$, $\langle F[\overline{\phi}(\mathbf{x},t)]\rangle$, using the expression

$$\langle F[\overline{\phi}(\mathbf{x},t)]\rangle = \int d\varphi F(\varphi)\rho_1[\varphi(\mathbf{x},t)]. \tag{13}$$

Eq. (13) explains the sense in which the transition from quantum to classical behaviour takes place here: though the scalar field $\overline{\phi}$ remains a quantum operator formally, we can introduce an auxiliary classical stochastic scalar field $\varphi(\mathbf{x},t)$ with the PDF determined from Eq. (12) that has the same expectation values for all observables with excellent accuracy (the error is due to the decaying mode of $\overline{\phi}$ that is exponentially small, $\propto \exp(-3Ht)$).

If $H$ is not exactly constant, but slowly varying, it is more natural to write the Fokker-Planck equation in terms of the independent variable $\ln a \equiv \int H(t)\, dt$ (see the discussion in [12]). Then the PDF $P_{\rho}(\varphi, \ln a)$ introduced as in [15] which refers to the "whole Universe" reduces to $\rho_1(\varphi, \ln a)$ after being normalized to unity according to the prescription given in [16]. Also, the distribution of domains $N(\varphi, \ln a)$ used in [17] is equal to $a^3\rho_1(\varphi, \ln a)$, if one takes $\epsilon H = const$ as pointed above. Clearly, in our case there is no difference between $Ht$ and $\ln a$ at all.

In general, the solution of (12) can be written in the form

$$\rho_1(\varphi,t) = \exp\left(-\frac{4\pi^2 V(\varphi)}{3H^4}\right)\sum_{n=0}^{\infty}a_n\Phi_n(\varphi)e^{-\Lambda_n(t-t_0)}, \tag{14}$$

where $\Phi_n(\varphi)$ is the complete orthonormal set of eigenfunctions of the Schrödinger-type equation

$$\left[-\frac{1}{2}\frac{\partial^2}{\partial\varphi^2}+W(\varphi)\right]\Phi_n(\varphi)=\frac{4\pi^2\Lambda_n}{H^3}\Phi_n(\varphi), \tag{15}$$

with the effective potential

$$W(\varphi)=\frac{1}{2}\left[v'(\varphi)^2-v''(\varphi)\right], \qquad v(\varphi)\equiv\frac{4\pi^2}{3H^4}V(\varphi). \tag{16}$$

The coefficients $a_n$ are given by an initial condition of $\rho_1$ at $t=t_0$ as

$$a_n=\int d\varphi \rho_1(\varphi,t)e^{v(\varphi)}\Phi_n(\varphi). \tag{17}$$

Because the left-hand-side of (15) can be recast in the form

$$\frac{1}{2}\left(-\frac{\partial}{\partial\varphi}+v'(\varphi)\right)\left(-\frac{\partial}{\partial\varphi}+v'(\varphi)\right)^\dagger\Phi_n(\varphi),$$

the eigenvalues $\Lambda_n$'s are nonnegative. If

$$N\equiv\int_{-\infty}^\infty e^{-2v(\varphi)}d\varphi, \tag{18}$$

is finite, we find $\Lambda_0=0$ and the corresponding eigenfunction is given by

$$\Phi_0(\varphi)=N^{-\frac{1}{2}}e^{-v(\varphi)}. \tag{19}$$

Then the normalization condition $\int_{-\infty}^\infty d\varphi\, \rho_1(\varphi,t)=1$ results in $a_0=N^{-\frac{1}{2}}$, so that (14) reads

$$\rho_1(\varphi,t)=\rho_{1q}(\varphi)+e^{-v(\varphi)}\sum_{n=1}^\infty a_n\Phi_n(\varphi)e^{-\Lambda_n(t-t_0)}. \tag{20}$$

Therefore, any solution (14) asymptotically approaches the static equilibrium solution

$$\rho_{1q}(\varphi)\equiv N^{-1}\exp\left(-\frac{8\pi^2}{3H^4}V(\varphi)\right)=N^{-1}e^{-2v(\varphi)} \tag{21}$$

This PDF is significantly non-Gaussian in the presence of self-interaction. In particular, if $V(\varphi)=\frac{\lambda}{4}\varphi^4$, then

$$\rho_{1q}(\varphi)=\left(\frac{32\pi^2\lambda}{3}\right)^{\frac{1}{4}}\frac{1}{\Gamma\left(\frac{1}{4}\right)H}\exp\left(-\frac{2\pi^2\lambda\varphi^4}{3H^4}\right). \tag{22}$$

One of distinctive non-Gaussian properties of the PDF (22) is its negative kurtosis $K=\frac{\langle\varphi^4\rangle}{\langle\varphi^2\rangle^2}-3\approx -0.812$. As for the dispersion of $\varphi$, it gives

$$\langle\varphi^2\rangle=\sqrt{\frac{3}{2\pi^2}}\frac{\Gamma(\frac{3}{4})}{\Gamma(\frac{1}{4})}\frac{H^2}{\sqrt\lambda}\approx 0.132\,\frac{H^2}{\sqrt\lambda}. \tag{23}$$

Comparing it with the Hartree-Fock result for this quantity presented at the end of the previous section, we see that the latter is $\approx 15\%$ less—not a bad result for the Hartree-Fock approximation. Certainly, it cannot be used to calculate such things as the kurtosis at all.

## IV. EQUILIBRIUM TEMPORAL AUTOCORRELATION FUNCTION IN THE STOCHASTIC APPROACH

Of course, the one-point PDF is not enough to obtain the whole picture of the scalar field distribution in the de Sitter space-time. We need to know all two-point and higher-point correlation functions. We shall show now that no new equations are necessary for this aim and that all these functions can be constructed from various solutions of the Fokker-Planck equation (12) (note that we are speaking about functions having anomalously large infrared part only). Let us begin with the temporal autocorrelation function of the scalar field.

The autocorrelation function (or the temporal two-point function) $G(t_1, t_2) \equiv \langle \overline{\phi}(\mathbf{x}, t_1) \overline{\phi}(\mathbf{x}, t_2) \rangle$ can be calculated from the two-point PDF at equal space points $\rho_2[\overline{\phi}(\mathbf{x}, t_1) = \varphi_1, \overline{\phi}(\mathbf{x}, t_2) = \varphi_2] \equiv \rho_2[\varphi_1(\mathbf{x}, t_1), \varphi_2(\mathbf{x}, t_2)]$ as

$$G(t_1, t_2) = \int d\varphi_1 d\varphi_2 \varphi_1 \varphi_2 \rho_2[\varphi_1(\mathbf{x}, t_1), \varphi_2(\mathbf{x}, t_2)]. \tag{24}$$

Here $\rho_2[\varphi_1(\mathbf{x}, t_1), \varphi_2(\mathbf{x}, t_2)]$ is given by

$$\rho_2[\varphi_1(\mathbf{x}, t_1), \varphi_2(\mathbf{x}, t_2)] = \Pi[\varphi_1(\mathbf{x}, t_1)|\varphi_2(\mathbf{x}, t_2)]\rho_1[\varphi_2(\mathbf{x}, t_2)]\theta(t_1 - t_2) \\ + \Pi[\varphi_2(\mathbf{x}, t_2)|\varphi_1(\mathbf{x}, t_1)]\rho_1[\varphi_1(\mathbf{x}, t_1)]\theta(t_2 - t_1), \tag{25}$$

where $\Pi[\varphi_1(\mathbf{x}, t_1)|\varphi_2(\mathbf{x}, t_2)]$ is the conditional probability to find $\overline{\phi}(\mathbf{x}, t_1) = \varphi_1$ provided that $\overline{\phi}(\mathbf{x}, t_2) = \varphi_2$. It satisfies the Fokker-Planck equation (12) with respect to both pairs of its arguments $\varphi_1, t_1$ and $\varphi_2, t_2$ [18]:

$$\frac{\partial \Pi}{\partial t_1}[\varphi_1(\mathbf{x}, t_1)|\varphi_2(\mathbf{x}, t_2)] = \Gamma_{\varphi_1} \Pi[\varphi_1(\mathbf{x}, t_1)|\varphi_2(\mathbf{x}, t_2)] \tag{26}$$

and

$$\frac{\partial \Pi}{\partial t_2}[\varphi_1(\mathbf{x}, t_1)|\varphi_2(\mathbf{x}, t_2)] = \Gamma_{\varphi_2} \Pi[\varphi_1(\mathbf{x}, t_1)|\varphi_2(\mathbf{x}, t_2)], \tag{27}$$

with the initial condition

$$\Pi[\varphi_1(\mathbf{x}, t_1)|\varphi_2(\mathbf{x}, t_1)] = \delta(\varphi_1 - \varphi_2). \tag{28}$$

Since $\Gamma_{\varphi_i}$ is independent of time, we find $\Pi[\varphi_1(\mathbf{x}, t_1)|\varphi_2(\mathbf{x}, t_2)]$ depends on time only through $|t_1 - t_2|$ using (26)–(28). For $t_1 > t_2$, $\rho_2[\varphi_1(\mathbf{x}, t_1), \varphi_2(\mathbf{x}, t_2)]$ itself satisfies

$$\frac{\partial \rho_2}{\partial t_1}[\varphi_1(\mathbf{x}, t_1), \varphi_2(\mathbf{x}, t_2)] = \Gamma_{\varphi_1} \rho_2[\varphi_1(\mathbf{x}, t_1), \varphi_2(\mathbf{x}, t_2)]. \tag{29}$$

For a de Sitter invariant state, $\langle \overline{\phi}(\mathbf{x}, t_1) \overline{\phi}(\mathbf{x}, t_2) \rangle$ depends only on $|t_1 - t_2|$, so that

$$G(t_1, t_2) = G(|t_1 - t_2|) = \langle \overline{\phi}^2 \rangle \xi(|t_1 - t_2|), \quad \xi(0) = 1. \tag{30}$$

This is realized in (24) if and only if $\rho_1[\varphi(\mathbf{x}, t)]$ is time-independent. In other words, in the language of stochastic inflation, the absence of a de Sitter invariant state for a free massless minimally-coupled scalar field is due to the non-existence of any static solution to (12) with $V(\phi) = 0$. On the contrary, one may expect to recover the de Sitter invariance of the two-point correlation function by adding the self-interaction $V(\phi) = \frac{\lambda}{4}\phi^4$ for which we have found the static equilibrium solution (22).

## A. Massive non-interacting case

Before proceeding to the explicit calculation of the autocorrelation function for the massless $\frac{\lambda}{4}\phi^4$ theory, let us demonstrate that we can obtain the familiar formula for the Green function of a free massive scalar field by taking $V(\phi) = \frac{1}{2}m^2\phi^2$ with $0 < m^2 \ll H^2$. Multiplying (29) by $\varphi_1 \varphi_2$ and integrating over $\varphi_1$ and $\varphi_2$, we find

$$\frac{\partial}{\partial t_1}\langle \overline{\phi}(t_1)\overline{\phi}(t_2)\rangle = -\frac{1}{3H}\langle V'[\overline{\phi}(t_1)]\overline{\phi}(t_2)\rangle = -\frac{m^2}{3H}\langle \overline{\phi}(t_1)\overline{\phi}(t_2)\rangle. \tag{31}$$

Since the equilibrium value of $\langle \overline{\phi}^2(t)\rangle$ can be calculated as

$$\langle \overline{\phi}^2(t)\rangle = \int d\varphi \varphi^2 \rho_{1q}(\varphi) = \frac{3H^4}{8\pi^2 m^2}, \tag{32}$$

Eq. (31) is solved to yield

$$\langle \overline{\phi}(t_1)\overline{\phi}(t_2)\rangle = \frac{3H^4}{8\pi^2 m^2} \exp\left(-\frac{m^2}{3H}|t_1 - t_2|\right). \tag{33}$$

This has to be compared with the two-point Green function in the de Sitter invariant vacuum calculated by methods of the field theory [1–3], namely,

$$\langle \phi(\mathbf{x_1}, t_1)\phi(\mathbf{x_2}, t_2)\rangle = \frac{H^2(1-c)(2-c)}{16\pi \sin \pi(1-c)} F\left(c, 3-c, 2; \frac{1+z}{2}\right); \quad c = \frac{3}{2} - \sqrt{\frac{9}{4} - \frac{m^2}{H^2}} \cong \frac{m^2}{3H^2}, \tag{34}$$

where $F$ is the hyper-geometric function and $z = z(x_1, x_2)$ is the de Sitter invariant function of $(\mathbf{x_1}, t_1)$ and $(\mathbf{x_2}, t_2)$ defined by

$$z(x_1, x_2) = \cosh H(t_1 - t_2) - \frac{H^2}{2} a_0^2 e^{Ht_1 + Ht_2} |\mathbf{x_1} - \mathbf{x_2}|^2. \tag{35}$$

It is related to the geodesic interval $s$ between $(\mathbf{x_1}, t_1)$ and $(\mathbf{x_2}, t_2)$ by the formula $z = 1 + H^2 s^2/2$. For $\mathbf{x_1} = \mathbf{x_2}$ and $|t_1 - t_2| \gg H^{-1}$, the leading term of (34) reads

$$\langle \phi(t_1)\phi(t_2)\rangle \simeq \frac{3H^4}{8\pi^2 m^2} \exp\left(-\frac{m^2}{3H}|t_1 - t_2|\right), \tag{36}$$

in agreement with (33). Since the autocorrelation function $G(|t|)$ is the integral over $\varphi_1$ and $\varphi_2$ of some solution of the Fokker-Planck equation (12) (see (24)), it itself may be expanded using the eigenvalues of Eq. (15):

$$G(|t|) = \sum_{n=1}^{\infty} b_n e^{-\Lambda_n |t|}, \quad \sum_{n=1}^{\infty} b_n = \langle \overline{\phi}^2 \rangle \tag{37}$$

(the coefficient $b_0 = 0$ due to the reflection symmetry). If $\lambda = 0$, then $\Lambda_n = m^2 n/3H$, $n = 0, 1, 2...$. We see that in the absence of self-interaction the autocorrelation function contains a contribution from the only one eigenvalue $\Lambda_1$. Let us introduce the correlation time $t_c$ using the definition $\xi(t_c) = 0.5$. Then $t_c = \ln 2 \, \Lambda_1^{-1} = 3 \ln 2 \, Hm^{-2} \gg H^{-1}$ for $\lambda = 0$.

## B. Massless interacting case

Now we turn to the massless minimally-coupled $\lambda\phi^4/4$ theory for which Eq. (31) is not readily soluble, and we must deal with the two-point PDF directly. Let us define the following dimensionless quantities

$$\tau \equiv \sqrt{\frac{\lambda}{24\pi^2}} Ht, \quad \tilde{\varphi} \equiv \left(\frac{8\pi^2\lambda}{3}\right)^{\frac{1}{4}} \frac{\varphi}{H} \equiv Q\varphi, \quad \tilde{\rho}_1 \equiv Q^{-1}\rho_1, \quad \tilde{\Pi} \equiv Q^{-1}\Pi, \quad \tilde{\rho}_2 \equiv Q^{-2}\rho_2 \quad (38)$$

with which the Fokker-Planck equation reads

$$\frac{\partial \tilde{\rho}}{\partial \tau} = \frac{\partial^2 \tilde{\rho}}{\partial \tilde{\varphi}^2} + \frac{\partial}{\partial \tilde{\varphi}}(\tilde{\varphi}^3 \tilde{\rho}) \equiv \tilde{\Gamma}_{\tilde{\varphi}} \tilde{\rho}. \quad (39)$$

The corresponding dimensionless effective potential $\tilde{W}(\tilde{\varphi}) = Q^{-2}W(\varphi)$, see (16), is depicted in Fig. 1.

Then the dimensionless autocorrelation function $g(\tau) \equiv \langle \tilde{\varphi}(\tau_0 + \tau)\tilde{\varphi}(\tau_0) \rangle = Q^2 G(t)$ is given by

$$g(\tau) = \int d\tilde{\varphi}_1 d\tilde{\varphi}_2 \tilde{\varphi}_1 \tilde{\Pi}[\tilde{\varphi}_1(\tau_0 + \tau)|\tilde{\varphi}_2(\tau_0)]\tilde{\rho}_{1\text{eq}}(\tilde{\varphi}_2)\tilde{\varphi}_2. \quad (40)$$

Using the Fokker-Planck equation for $\tilde{\Pi}$ and the initial condition $\tilde{\Pi}[\tilde{\varphi}_1(\tau_0)|\tilde{\varphi}_2(\tau_0)] = \delta(\tilde{\varphi}_1 - \tilde{\varphi}_2)$, we find

$$\left.\frac{d^n g(\tau)}{d\tau^n}\right|_{\tau=0} = \int d\tilde{\varphi}_1 d\tilde{\varphi}_2 \tilde{\varphi}_1 \tilde{\Gamma}^n_{\tilde{\varphi}_1} \tilde{\Pi}[\tilde{\varphi}_1(\tau_0)|\tilde{\varphi}_2(\tau_0)]\tilde{\rho}_{1\text{eq}}(\tilde{\varphi}_2)\tilde{\varphi}_2. = \int d\tilde{\varphi}_1 \tilde{\varphi}_1 \tilde{\Gamma}^n_{\tilde{\varphi}_1} \tilde{\rho}_{1\text{eq}}(\tilde{\varphi}_1)\tilde{\varphi}_1, \quad (41)$$

with which we can express $g(\tau)$ formally as a Taylor series,

$$g(\tau) = \sum_{n=0}^{\infty} \frac{\tau^n}{n!} \frac{d^n g(0)}{d\tau^n}. \quad (42)$$

All quantities (41) can be calculated analytically:

$$g(0) \equiv \langle \tilde{\varphi}^2 \rangle = \frac{2\Gamma(\frac{3}{4})}{\Gamma(\frac{1}{4})} \approx 0.6760, \quad \frac{dg(0)}{d\tau} = -1, \quad \frac{d^2 g(0)}{d\tau^2} = 3g(0),$$

$$\frac{d^3 g(0)}{d\tau^3} = -9, \quad \frac{d^4 g(0)}{d\tau^4} = 117g(0) \ldots \quad (43)$$

On the other hand, the series (42) is not useful for the calculation of $g(\tau)$ for a finite $\tau$ since it is only an asymptotic one and it has the zero radius of convergence. So, to find the full autocorrelation function we resort to a numerical calculation. To be specific, we numerically solve the Fokker-Planck equation (39) for a new function $\Xi$,

$$\frac{\partial \Xi(\tilde{\varphi}_1, \tau)}{\partial \tau} = \tilde{\Gamma}_{\tilde{\varphi}_1} \Xi(\tilde{\varphi}_1, \tau), \quad \Xi(\tilde{\varphi}_1, \tau) \equiv \int d\tilde{\varphi}_2 \tilde{\Pi}[\tilde{\varphi}_1(\tau)|\tilde{\varphi}_2(0)]\tilde{\rho}_{1\text{eq}}(\tilde{\varphi}_2)\tilde{\varphi}_2, \quad (44)$$

with the initial condition

$$\Xi(\tilde{\varphi}_1, 0) = \tilde{\rho}_{1\text{eq}}(\tilde{\varphi}_1)\tilde{\varphi}_1. \tag{45}$$

Then we can find $g(\tau)$ from the integral

$$g(\tau) = \int d\tilde{\varphi}_1 \tilde{\varphi}_1 \Xi(\tilde{\varphi}_1, \tau). \tag{46}$$

The result is shown in Fig. 2. We find that $g(\tau)$ asymptotically varies as $g(\tau) \propto e^{-1.36859\tau}$. As expected, the numerical factor in the denominator is related to the lowest nonvanishing eigenvalue of (15):

$$\Lambda_1 \approx 1.36859 \sqrt{\frac{\lambda}{24\pi^2}} H. \tag{47}$$

Of course, $g(\tau)$ is not purely exponential as in the $\lambda = 0$ case considered above, eigenvalues $\Lambda_n$ with $n \geq 2$ make some contribution to its expansion of the form (37). However, this additional contribution is small already at $\tau = 0$ (e.g., $|g^{-1}(0)\frac{dg(0)}{d\tau}|$ is only 8% more than its asymptotical value 1.36859 for $\tau \gg 1$), and it dies quickly because higher eigenvalues are significantly larger (in particular, $\Lambda_2 \approx 4.4537\sqrt{\frac{\lambda}{24\pi^2}}H$). As a result, $g(\tau)$ becomes purely exponential with the error less than 1% for $\tau > 0.3$. $\xi(\tau) = 0.5$ at $\tau \approx 0.495$, so the correlation time is $t_c \approx 7.62/H\sqrt{\lambda}$.

### C. Interacting case with two potential wells

Another interesting case takes place when $m^2 < 0$ and the potential has the form of a double well:

$$V(\phi) = \frac{\lambda}{4}(\phi^2 - \phi_0^2)^2, \quad \phi_0^2 = \frac{|m^2|}{\lambda}. \tag{48}$$

In this model, the scalar field fluctuates around $\phi = \phi_0$ in some domains and around $\phi = -\phi_0$ in others. The correlation time, which is characterized by the lowest nonzero eigenvalue, $\Lambda_1$, of eq. (15), is much larger than in the massless $\lambda\phi^4$ theory. Let us assume $\sqrt{\lambda}H^2 \ll |m^2| \ll H^2$ to demonstrate it. The effective potential in the Schrödinger equation (15) for the double-well potential (48) (made dimensionless using the quantities (38)) has the form:

$$\tilde{W}(\tilde{\varphi}) = Q^{-2}W(\varphi) = \frac{1}{8}\tilde{\varphi}^2(\tilde{\varphi}^2 + \gamma)^2 - \frac{1}{4}(3\tilde{\varphi}^2 + \gamma),$$

$$\gamma = \frac{m^2 Q^2}{\lambda} = \sqrt{\frac{8\pi^2}{3\lambda}}\frac{m^2}{H^2} < 0. \tag{49}$$

It is depicted in Fig. 1 for $\gamma = -4$.

If the wells are sufficiently far from each other, then each energy level existing in one well in the absence of the other (including the lowest one in which we are interested in) becomes splitted into two with an energy difference between them being exponentially small. So, the eigenfunctions corresponding to $\Lambda_0 = 0$ and $\Lambda_1$ may be represented, respectively, as

$$\Phi_0(\varphi) = \frac{1}{\sqrt{2}} \left[ \Phi_*(\varphi) + \Phi_*(-\varphi) \right], \tag{50}$$

$$\Phi_1(\varphi) = \frac{1}{\sqrt{2}} \left[ \Phi_*(\varphi) - \Phi_*(-\varphi) \right], \tag{51}$$

where $\Phi_*(\varphi)$ is the normalized ground-state function for the single well localized around $\varphi = \varphi_b \approx \phi_0$ with $\varphi_b$ minimizing the potential $W(\varphi)$. The problem we want to solve is close to that investigated in [19] (§ 50, Problem 3) in the WKB approximation. However, if we pretend to obtain not only the correct exponent, but the coefficient of the exponential, too, we may not directly apply the formula given there because the WKB approximation does not work for the lowest energy level generally. So, instead of the WKB approximation, we shall use the perturbation theory with respect to the exponentially small quantity $\varepsilon_1 = 4\pi^2 \Lambda_1/H^3$.

Two independent solutions of Eq. (15) in the static case $\Lambda_n = 0$ are:

$$\Phi_{st}^{(0)} = e^{-v(\varphi)}, \quad \Phi_{st}^{(1)} = e^{-v(\varphi)} \int^{\varphi} e^{2v(\varphi_1)} \, d\varphi_1. \tag{52}$$

The former function is actually proportional to that in Eq. (19). The corresponding static Green function of Eq. (15) is:

$$G_{st}(\varphi, \varphi_1) = \Phi_{st}^{(1)}(\varphi)\Phi_{st}^{(0)}(\varphi_1) - \Phi_{st}^{(1)}(\varphi_1)\Phi_{st}^{(0)}(\varphi). \tag{53}$$

Thus, the unnormalized eigenfunction for first non-zero energy level has the following form in the region $\varphi \gg H^2|m|^{-1}$ (where the correction to the unperturbed wave function is small):

$$\Phi_1(\varphi) = e^{-v(\varphi)} - 2\varepsilon_1 \int_{\infty}^{\varphi} G_{st}(\varphi, \varphi_1) e^{-v(\varphi_1)} \, d\varphi_1$$
$$= e^{-v(\varphi)} \left( 1 - 2\varepsilon_1 \int_{\phi_0}^{\varphi} e^{2v(\varphi_1)} \, d\varphi_1 \int_{\infty}^{\varphi} e^{-2v(\varphi_2)} \, d\varphi_2 + 2\varepsilon_1 \int_{\infty}^{\varphi} e^{-2v(\varphi_1)} \, d\varphi_1 \int_{\phi_0}^{\varphi_1} e^{2v(\varphi_2)} \, d\varphi_2 \right). \tag{54}$$

Here the lower integration limit in the formula (52) for $\Phi_{st}^{(1)}$ is chosen to be $\varphi = \phi_0$, i.e., to lie in the minimum of the potential $V(\phi)$. However, any other value satisfying the condition $\varphi \gg H^2|m|^{-1}$ may be used instead of it within the accuracy required to obtain a correct asymptotic value of $\Lambda_1$.

For $\varphi \ll \phi_0$, the expression (54) takes the form:

$$\Phi_1(\varphi) \approx e^{-v(\varphi)} \left( 1 + \varepsilon_1 N \int_{\phi_0}^{\varphi} e^{2v(\varphi_1)} \, d\varphi_1 \right). \tag{55}$$

The normalization coefficient $N$ is defined in Eq. (19). In our case it is equal to:

$$N = \int_{-\infty}^{\infty} e^{-2v(\varphi)} \, d\varphi = \sqrt{\frac{3}{8\pi^2} \frac{H^2}{|m|}} \tag{56}$$

(the main contribution is made by vicinities of the points $\pm\phi_0$). Now $\Phi_1(\varphi)$ in Eq. (55) is represented as a linear combination of the two exact unperturbed solutions (52). Therefore, we may use this expression even if the second term inside the round brackets in its right-hand side becomes comparable to the first term.

On the other hand, $\Phi_1(\varphi)$ should be an odd function of $\varphi$, see (51). Therefore, $\Phi_1(0) = 0$. This gives the required equation for $\varepsilon_1$:

$$\varepsilon_1 N \int_0^{\phi_0} e^{2v(\varphi_1)} \, d\varphi_1 = 1. \tag{57}$$

We, therefore, find

$$\Lambda_1 = \frac{\sqrt{2} m^2}{3\pi H} \exp\left(-\frac{2\pi^2 m^4}{3\lambda H^4}\right). \tag{58}$$

The correlation time is $t_c \approx \ln 2 \, \Lambda_1^{-1}$ that is exponentially larger than in the case of the massless $\lambda \phi^4$ theory.

Similar to the massive non-interacting case, higher eigenvalues are given by the expression

$$\Lambda_{2n,\ 2n+1} \approx \frac{2|m^2|n}{3H}, \quad n = 1, 2... \tag{59}$$

neglecting exponentially small splitting (the effective mass at the bottom of each well is equal to $\sqrt{2}\,|m|$). Using the spectral representation (37) and the formula (33), we arrive at the following expression for the autocorrelation function:

$$G(t) \approx \frac{|m^2|}{\lambda} + \frac{3H^4}{16\pi^2 |m^2|} \exp\left(-\frac{2|m^2| t}{3H}\right), \quad t \ll \Lambda_1^{-1}, \tag{60}$$

$$\approx \frac{|m^2|}{\lambda} \exp(-\Lambda_1 t), \quad t \gg \frac{H}{|m^2|}, \tag{61}$$

Eq. (60) shows the change in $G(t)$ due to decay of correlations inside one domain, while Eq. (61) describes how the correlations disappear completely at very large times due to quantum jumps between different domains.

Note that the exponent in Eq. (58) is nothing but the action for the Hawking-Moss (or de Sitter) instanton [20]. As discussed in [12], the physical sense of this instanton in the case under consideration becomes more transparent if one writes it in the static, "thermal" form:

$$ds^2 = (1 - H^2 r^2)\, dt_{st}^2 + (1 - H^2 r^2)^{-1}\, dr^2 + r^2\, d\Omega^2,$$
$$\phi = 0, \tag{62}$$

where $t_{st}$ is periodic with the period $2\pi H^{-1}$. Then it describes the "thermal" nucleation of a bubble of the phase with $\phi = -\phi_0$ inside the phase $\phi = \phi_0$ (or vice versa) through a quantum jump over the potential barrier which maximum is located at $\phi = 0$. The size of the bubble is equal to the size of the de Sitter event horizon $H^{-1}$.

However, deriving Eqs. (58,61) we have got more—we have obtained the correct coefficient of the exponential, too, that is equivalent to the calculation of a one-loop correction to the instanton contribution to $G(t)$. This shows that not only does the stochastic approach go beyond the Hartree-Fock approximation, but it is able to reproduce results obtained using the instanton approach in the path integral formalism.

## D. General case

Here we present formulas valid for the case of an arbitrary potential $V(\phi)$. Using the expressions (24–29) and repeating the calculation for the massless case presented above, we arrive at the following representation for the autocorrelation function in the equilibrium state:

$$G(t) = \int \varphi \Xi(\varphi, t) \, d\varphi, \tag{63}$$

where the function

$$\Xi(\varphi, t) \equiv \int \varphi_1 \rho_{1eq}(\varphi_1) \Pi\left[\varphi(t) | \varphi_1(0)\right] d\varphi_1 \tag{64}$$

satisfies the Fokker-Planck equation (12) with the initial condition

$$\Xi(\varphi, 0) = \varphi \rho_{1eq}(\varphi). \tag{65}$$

It is straightforward to generalize these formulas to the autocorrelation function of an arbitrary function $F(\varphi)$. The result is that in order to obtain the corresponding expressions for the correlation function

$$G_F(|t_1 - t_2|) \equiv \langle F[\overline{\phi}(\mathbf{x}, t_1)] F[\overline{\phi}(\mathbf{x}, t_2)] \rangle, \tag{66}$$

one has to substitute the multiplier $\varphi$ in the right-hand sides of Eqs. (63-65) by $F(\varphi)$. In other words, one has to solve the same Fokker-Planck equation with a different initial condition.

Also useful is the spectral representation of $G(t)$. For the initial condition (65), the coefficients (17) of expansion of $\Xi(\varphi, t)$ into the complete set of eigenfunctions of Eq. (15) (which we denote by $A_n$ for this concrete case) have the form:

$$A_n = \int \varphi \rho_{1eq}(\varphi) e^{v(\varphi)} \Phi_n(\varphi) \, d\varphi = N^{-1} \int \varphi e^{-v(\varphi)} \Phi_n(\varphi) \, d\varphi, \tag{67}$$

where the normalization coefficient $N$ is given in (19) and $v$ is defined in (16). Note that $A_0 = 0$ if $V(-\varphi) = V(\varphi)$. Then, using Eqs. (14) and (63), we arrive at the formula:

$$G(t) = \sum A_n e^{-\Lambda_n t} \int \varphi e^{-v(\varphi)} \Phi_n(\varphi) \, d\varphi = N \sum A_n^2 e^{-\Lambda_n t}. \tag{68}$$

The relations (43) for the massless theory may be generalized, too. After differentiation of Eq. (63) and application of the Fokker-Planck equation, successive time derivatives of $G(t)$ at $t = 0$ follow:

$$\frac{dG(0)}{dt} = -\frac{H^3}{8\pi^2}, \tag{69}$$

$$\frac{d^2G(0)}{dt^2} = \frac{H^2}{24\pi^2 N} \int \frac{d^2V(\varphi)}{d\varphi^2} e^{-2v(\varphi)} \, d\varphi = \frac{H^2}{24\pi^2} \left(3\lambda G(0) + m^2\right) \tag{70}$$

and so on (in the last equality in Eq. (70), we have inserted $V(\varphi) = \frac{\lambda \varphi^4}{4} + \frac{m^2 \varphi^2}{2}$). Another quantity for which an analytical expression may be obtained is $I = \int_0^\infty G(t) \, dt$. Applying the method described in [11,12], we get

$$I = \frac{32\pi^2}{H^3 N} \int_0^\infty e^{2v(\varphi)} \, d\varphi \left( \int_\varphi^\infty \varphi_1 e^{-2v(\varphi_1)} \, d\varphi_1 \right)^2 \tag{71}$$

(the expression is written for the case of a symmetric potential $V(-\varphi) = V(\varphi)$ for simplicity). Eq. (71) may be used, e.g., to derive the formula (58) in a completely different way.

A note should be made about a region of applicability of the expression for $G(t)$. According to the derivation of the Fokker-Planck equation in Sec. 3, we consider time intervals larger than $H^{-1}|\ln \epsilon| \gg H^{-1}$. So, strictly speaking, all formulas for $G(t)$ are valid under this condition, too. However, $G(t)$ is practically constant and equal to its value at zero lag $G(0) = \langle \phi^2 \rangle$ for $Ht \ll \min(H^2|m^{-2}|, \lambda^{-1/2})$ already. The expressions (69,70) for the time derivatives at zero lag should be understood in the same sense, i.e., actually they refer to the argument lying in the interval $H^{-1} \ll t \ll H^{-1} \min(H^2|m^{-2}|, \lambda^{-1/2})$.

If $\lambda$ remains fixed and $m^2$ decreases from positive to negative values, two main trends follow from the results presented above:
1) the autocorrelation function (in particular, $G(0)$) grows,
2) the correlation time $t_c$ grows very quickly.
To visualize these trends we made a numerical calculation of the dimensionless correlation function $g(\tau) \equiv Q^2 G(t)$ for the cases $\gamma = 1, 0, -1, -4$ using the dimensionless variables defined in (38,49). The results are depicted in Figs. 2 and 3. They confirm the expected behaviour. The case $\gamma = -4$ is already close to the asymptotic case considered in subsection C, in particular, the lowest non-zero eigenvalue $\Lambda_1$ differs by less than 10% from that given by Eq. (58).

## V. GENERAL EQUILIBRIUM TWO-POINT CORRELATION FUNCTION AND ITS DE SITTER INVARIANCE

Let us now consider the spatial correlation function

$$G(r,t) = \langle \overline{\phi}(\mathbf{x}_1, t) \overline{\phi}(\mathbf{x}_2, t) \rangle, \quad r = |\mathbf{x}_1 - \mathbf{x}_2|. \tag{72}$$

Repeating the same procedure to derive (12), one can also find the Fokker-Planck equation for the two-point PDF at equal time $\rho_2[\varphi_1(\mathbf{x}_1, t), \varphi_2(\mathbf{x}_2, t)]$ as

$$\frac{\partial \rho_2}{\partial t}[\varphi_1(\mathbf{x}_1, t), \varphi_2(\mathbf{x}_2, t)] = \Gamma_{\varphi_1} \rho_2 + \Gamma_{\varphi_2} \rho_2 + j_0(\epsilon a(t) H |\mathbf{x}_1 - \mathbf{x}_2|) \frac{H^3}{4\pi^2} \frac{\partial^2 \rho_2}{\partial \varphi_1 \partial \varphi_2}. \tag{73}$$

Since $j_0(z)$ is a rapidly oscillating function for $|z| \gtrsim 1$ and we are concerned with coarse-grained quantities, we can, and in fact we should, adopt the following approximation:

$$j_0(\epsilon a(t) H |\mathbf{x}_1 - \mathbf{x}_2|) \simeq \theta(1 - \epsilon a(t) H |\mathbf{x}_1 - \mathbf{x}_2|), \tag{74}$$

which is employed hereafter. The above replacement corresponds to the approximation that two points $\mathbf{x}_1$ and $\mathbf{x}_2$ with their proper distance $a(t)|\mathbf{x}_1 - \mathbf{x}_2| \equiv a(t)r < (\epsilon H)^{-1}$ have 100% correlation while those with larger separation are mutually independent at the time $t$.

In fact, one can readily see that

$$\rho_2(\varphi_1, \varphi_2) \equiv \rho_{1q}(\varphi_1)\delta(\varphi_1 - \varphi_2) \tag{75}$$

constitutes a static solution of (73) with $j_0$ replaced by unity. Therefore one can find the equilibrium two-point PDF with $a(t)r > (\epsilon H)^{-1}$ by solving

$$\frac{\partial \rho_2}{\partial t}[\varphi_1(\mathbf{x}_1, t), \varphi_2(\mathbf{x}_2, t)] = \Gamma_{\varphi_1}\rho_2 + \Gamma_{\varphi_2}\rho_2, \tag{76}$$

with the initial condition

$$\rho_2[\varphi_1(\mathbf{x}_1, t_r), \varphi_2(\mathbf{x}_2, t_r)] = \rho_{1q}(\varphi_1)\delta(\varphi_1 - \varphi_2), \quad Ht_r = -\ln(ra_0\epsilon H). \tag{77}$$

Now we turn to the calculation of the general two-point correlation function

$$G(r, t_1, t_2) = \langle \overline{\phi}(\mathbf{x}_1, t_1)\overline{\phi}(\mathbf{x}_2, t_2) \rangle. \tag{78}$$

Following essentially the same line of arguments, we arrive at the basic expression for the general two-point PDF in the stochastic approach:

$$\rho_2[\varphi_1(\mathbf{x}_1, t_1), \varphi_2(\mathbf{x}_2, t_2)] = \int \Pi[\varphi_1(\mathbf{x}_1, t_1)|\varphi_r(\mathbf{x}_1, t_r)]\Pi[\varphi_2(\mathbf{x}_2, t_2)|\varphi_r(\mathbf{x}_2, t_r)]\rho_{1q}(\varphi_r)\, d\varphi_r,$$
$$t_r = -H^{-1}\ln(ra_0\epsilon H), \quad r = |\mathbf{x}_1 - \mathbf{x}_2|, \tag{79}$$

where $\Pi[\varphi_1(\mathbf{x}, t_1)|\varphi_2(\mathbf{x}, t_2)]$ satisfies the Fokker-Planck equations (26,27) with respect to both its time arguments and the initial condition (28). Note also that the spatial points $\mathbf{x}_1$ and $\mathbf{x}_2$ were inside the same elementary averaging volume at the moment $t = t_r$, thus, they had the same one-point PDF at this time.

Of course, it is assumed here that $t_1, t_2 > t_r$. This means that the formula (79) refers to space-time points $(\mathbf{x}_1, t_1)$ and $(\mathbf{x}_2, t_2)$ lying outside each other's light cones. On the contrary, if these points may be connected by a time-like or null geodesics ($z \geq 1$ in Eq. (35)), then $\min(t_1, t_2) < t_r$ that corresponds to the points $(\mathbf{x}_1, \min(t_1, t_2))$ and $(\mathbf{x}_2, \min(t_1, t_2))$ lying inside one elementary averaging volume, i.e., *coinciding* in terms of the stochastic approach. Then the 2-point function $G(r, t_1, t_2)$ is equal to the autocorrelation function $G(|t_1 - t_2|)$ found in the previous section.

Returning to the case of a space-like separation between two points and using the spectral decomposition (14) for $\Pi[\varphi_1(\mathbf{x}, t_1)|\varphi_2(\mathbf{x}, t_2)]$ together with the definition (67) of $A_n$, we obtain

$$G(r, t_1, t_2) = \int d\varphi_1\, \varphi_1 \int d\varphi_2\, \varphi_2 \rho_2[\varphi_1(\mathbf{x}_1, t_1), \varphi_2(\mathbf{x}_2, t_2)]$$
$$= \iint d\varphi_1\, d\varphi_2\, \varphi_1\varphi_2 e^{-v(\varphi_1)}e^{-v(\varphi_2)} \int d\varphi_r e^{2v(\varphi_r)}\rho_{1q}(\varphi_r)$$
$$\times \sum_m \Phi_m(\varphi_1)\Phi_m(\varphi_r)e^{-\Lambda_m(t_1 - t_r)} \sum_n \Phi_n(\varphi_2)\Phi_n(\varphi_r)e^{-\Lambda_n(t_2 - t_r)}$$
$$= N\sum_m \sum_n \int d\varphi_r \Phi_m(\varphi_r)\Phi_n(\varphi_r)A_m A_n e^{-\Lambda_m(t_1 - t_r)}e^{-\Lambda_n(t_2 - t_r)}$$
$$= N\sum_n A_n^2 e^{-\Lambda_n(t_1 + t_2 - 2t_r)} = N\sum_n A_n^2 e^{-\Lambda_n(t_1 + t_2)} \exp\left(-2\Lambda_n H^{-1}\ln(ra_0\epsilon H)\right). \tag{80}$$

Now we may choose the small parameter $\epsilon$ of the coarse-graining (6) in such a way that $\epsilon^{-2\Lambda_n/H} \approx 1$ for all $\Lambda_n$ which make a significant contribution to $G(t)$. This is achieved if $\epsilon$ satisfies the inequalities

$$\exp\left(-\min(H^2|m^{-2}|,\lambda^{-1/2})\right) \ll \epsilon \ll 1. \tag{81}$$

As a result, we may omit $\epsilon$ from the last line in Eq. (80). Then, from the comparison of (80) with (68), the expression for the general two-point correlation function in the equilibrium state in terms of the autocorrelation function $G(t)$ found in the previous section follows:

$$G(r, t_1, t_2) = G\left(t_1 + t_2 + 2H^{-1}\ln(ra_0 H)\right). \tag{82}$$

Using the de Sitter invariant form $z(x_1, x_2)$ (35) in the regime of a strongly space-like separation between two points ($z < 0$, $|z| \gg 1$) to which the formula (79) refers, Eq. (82) can be rewritten in the form

$$G(r, t_1, t_2) = G\left(H^{-1}\ln|2z-1|\right) \tag{83}$$

that applies both to the cases of large space-like and time-like separations. As explained at the end of the previous section, we may use this expression even in the vicinity of the point $z = 1$ where the right-hand side of (83) is simply equal to $G(0) = \langle \phi^2 \rangle$. Therefore, we have proved the de Sitter invariance of the two-point correlation function.

In particular, the equal-time spatial correlation function (72)

$$G(r, t) \equiv G(r, t, t) = N \sum_n A_n^2 e^{-2\Lambda_n t} \exp\left(-2\Lambda_n H^{-1}\ln(ra_0 H)\right) = N \sum_n A_n^2 (RH)^{-2\Lambda_n/H} \tag{84}$$

appears to depend on the physical spatial distance $R \equiv ra_0 e^{Ht}$ only. Thus, a typical scale of spatial correlations in the equilibrium state, which defines the characteristic size of domains with positive and negative $\varphi$ and domain walls with $\varphi \sim 0$ between them, *does not* expand with the expansion of the $t = const$ hypersurface of the de Sitter space-time. Similar to the correlation time $t_c$, we can introduce the spatial physical correlation radius $R_c$ using the definition $G(R_c) = G(0)/2$. Then the de Sitter invariance property (82) gives the relation between $t_c$ and $R_c$:

$$R_c = H^{-1}\exp\left(\frac{Ht_c}{2}\right). \tag{85}$$

## VI. DISCUSSION

In the present paper, we have further developed the stochastic approach to inflation by constructing a method of calculation of the 2-point and higher-point correlation functions of a quantum scalar field in the de Sitter background in terms of solutions of the Fokker-Planck equation (12). This enables us to demonstrate the existence of the equilibrium state, which has the de Sitter invariance, for a self-interacting scalar field with a small mass term of an arbitrary sign. It is easy to see that this state is the unique attractor for any initial configuration. For example, the temporal two-point PDF $\rho_2[\varphi_1(\mathbf{x}, t_2 + t), \varphi_2(\mathbf{x}, t_2)]$ with an arbitrary initial distribution $\rho_1[\varphi_2(\mathbf{x}, t_2)]$ is obtained solving the Fokker-Planck equation (29) with the initial condition

$$\rho_2[\varphi_1(\mathbf{x}, t_2), \varphi_2(\mathbf{x}, t_2)] = \rho_1[\varphi_2(\mathbf{x}, t_2)]\delta(\varphi_1 - \varphi_2). \tag{86}$$

Since any solution $\rho_1[\varphi_2(\mathbf{x}, t_2)]$ approaches $\rho_{1q}(\varphi_2)$ as $t_2 \to \infty$, $\rho_2[\varphi_1(\mathbf{x}, t_2 + t), \varphi_2(\mathbf{x}, t_2)]$, as well as the autocorrelation function $G(t)$, approaches the equilibrium counterpart in this limit. Moreover, as long as we are interested in a fixed physical length scale, similar arguments hold for the spatial correlation function as well, and the asymptotic spatial correlation function does not depend on time even though the space (i.e., the hypersurface $t = const$ in the system of reference (3)) is expanding. It is evident that had we sticked to the conventional perturbative expansion around the free asymptotic state, we could not have obtained the above results.

Thus, any weakly interacting scalar field with a small mass eventually loses the memory of its initial state at the beginning of the de Sitter era as $t \to \infty$. The time scale for this to happen may be characterized by a relaxation time $t_{rel}$. Its value depends on the quantity involved. Generally, the relaxation time is of the order or less than the correlation time $t_c$, e.g., $t_{rel} \sim t_c \simeq \ln 2/\Lambda_1$ for the two-point PDF. On the other hand, repeating the calculations in Eqs. (68, 80) for an arbitrary non-equilibrium initial one-point PDF $\rho_1(\varphi, t_0)$, we arrive at the result that the relaxation time for the two-point correlation function $t_{rel} \sim \Lambda_2^{-1}$ for any reflection symmetric potential $V(-\phi) = V(\phi)$. So, in this case

$$t_{rel} \sim \min\left(H|m^{-2}|, H^{-1}\lambda^{-1/2}\right), \tag{87}$$

and $t_{rel}$ may be much less than $t_c$, cf. Eqs. (58) and (59).

Very important is the result (84) showing that the equal-time spatial correlation function depends on the physical radius $R$ only (and not on the comoving radius $r$). This has dramatic consequences for the general picture of fluctuations of the scalar field which are especially interesting in the case of a potential with two wells and $|m^2| \gg \sqrt{\lambda} H^2$ (Sec. 4, subsection C). In the latter case, the space is covered by clearly pronounced domains with $\varphi \approx \pm\phi_0$ with relatively thinner (but much thicker than $H^{-1}$) domain walls between them. Typical size of the domains in the equilibrium state is given by the spatial correlation radius (85). It is fantastically large because it has the form of a double exponent, see Eq. (58), but still it is finite. The size of the domain walls $R_{dw}$ is determined by $\Lambda_2^{-1}$ and contains single large exponent: $\ln(R_{dw}H) \sim H^2|m^{-2}|$. Both these physical sizes do not depend on time on average that corresponds to their typical comoving radius $r$ shrinking with time. Therefore, the expansion of the background physical space volume $\propto e^{3Ht}$ is not accompanied by the corresponding expansion of individual domains, instead of it, new domains separated by new domain walls are created with constant rate. This purely quantum (or stochastic) behaviour has to be contrasted with the deterministic classical expansion of inflationary domain walls well studied in the context of the new inflationary scenario (recently, renewed interest in this problem arose, especially, in connection with more complicated topological defects like strings and monopoles [21,22]). Though our calculations refer to the case of a test field in the stable de Sitter background, it is clear that one may expect a similar effect when a quasi-de Sitter background is produced by the scalar field itself, namely, that the classical monotonous quasi-exponential expansion of an individual domain wall at the first stage of inflation stops at the late asymptotic stage due to the loss of coherence produced by the quantum noise, and new domain walls are created instead of it. This will be considered in more details in another publication.

In this paper, we assumed the exact, "eternal" de Sitter background with $H = const$. In real inflationary models, $H$ is not exactly constant. The characteristic time of its change is $\tau_H = H/|\dot{H}| \gg H^{-1}$. Then the behaviour of a quantum self-interacting scalar field in such a background depends on the relation between $\tau_H$ and the relaxation time. If $t_{rel} \ll \tau_H$, the scalar field has enough time to reach the equilibrium state described in this paper for each successive $H(t)$. In the opposite case, evolution of the scalar field is significantly different and constitutes a separate problem.

## ACKNOWLEDGMENTS

A. S. is grateful to Profs. Y. Nagaoka and J. Yokoyama for their hospitality at the Yukawa Institute for Theoretical Physics, Kyoto University. A. S. was supported in part by the Russian Foundation for Basic Research, Project Code 93-02-3631, and by Russian Research Project "Cosmomicrophysics". J. Y. acknowledges support by the Japanese Grant-in-Aid for Scientific Research Fund of Ministry of Education, Science, and Culture, No. 06740216.

# Figure Captions

**Figure 1** Dimensionless effective potential $\tilde{W}[\check{\varphi}]$ for a quartic potential $V[\phi] = \frac{\lambda}{4}\phi^4$ (solid line), and that for a double well potential $V[\phi] = \frac{\lambda}{4}(\phi^2 - \phi_0^2)^2 = \frac{\lambda}{4}(\phi^2 + \gamma Q^{-2})^2$ with the dimentionless mass parameter $\gamma = -4$ (dashed line).

**Figure 2** Dimensionless autocorrelation function $g(\tau)$ of a scalar field with a potential $\frac{\lambda}{4}(\phi^2 + \gamma Q^{-2})^2$. Three curves represent the cases with $\gamma = -1, 0$, and 1, respectively, from the above.

**Figure 3** Same as Figure 2 but with $\gamma = -4$.

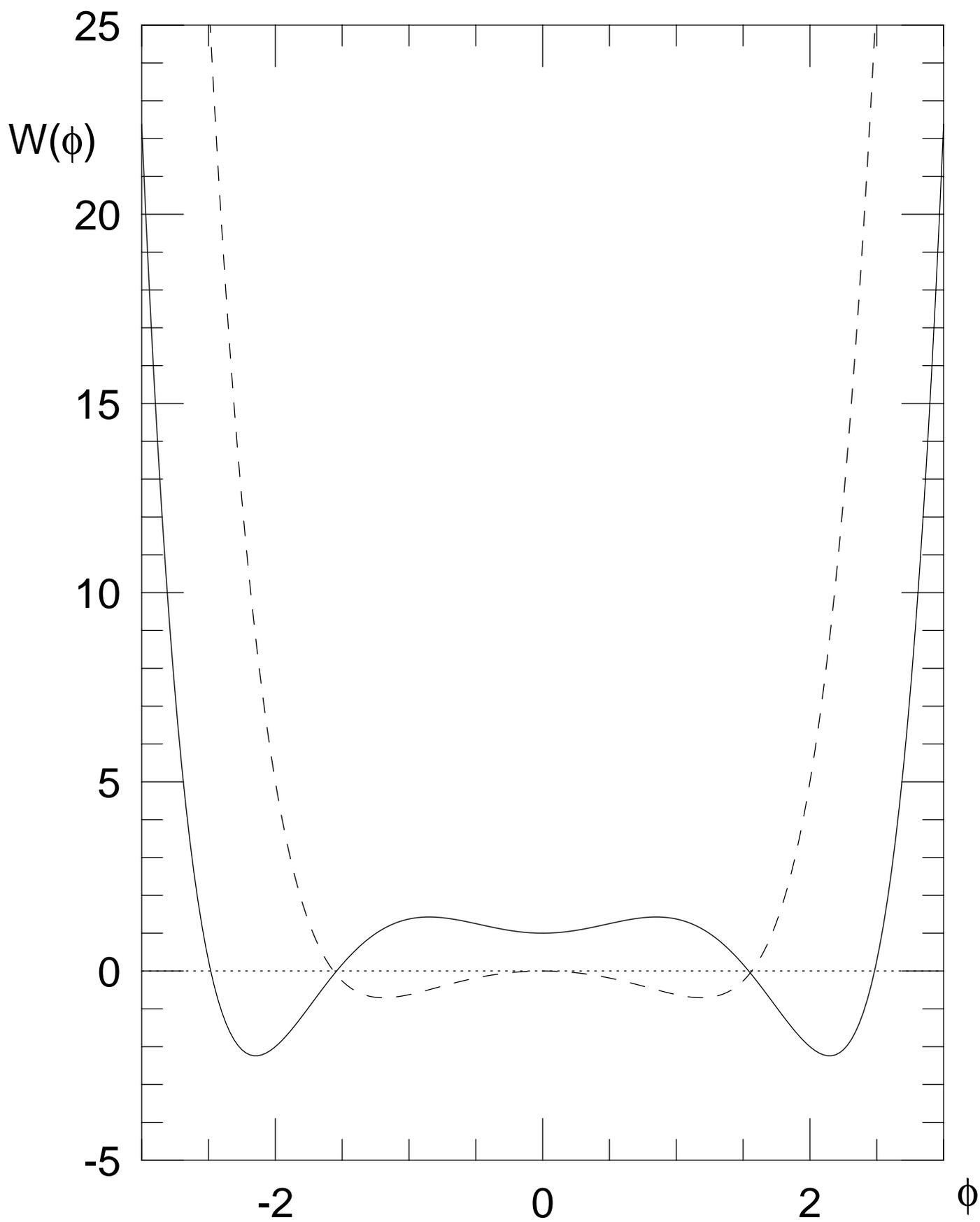

Figure 1

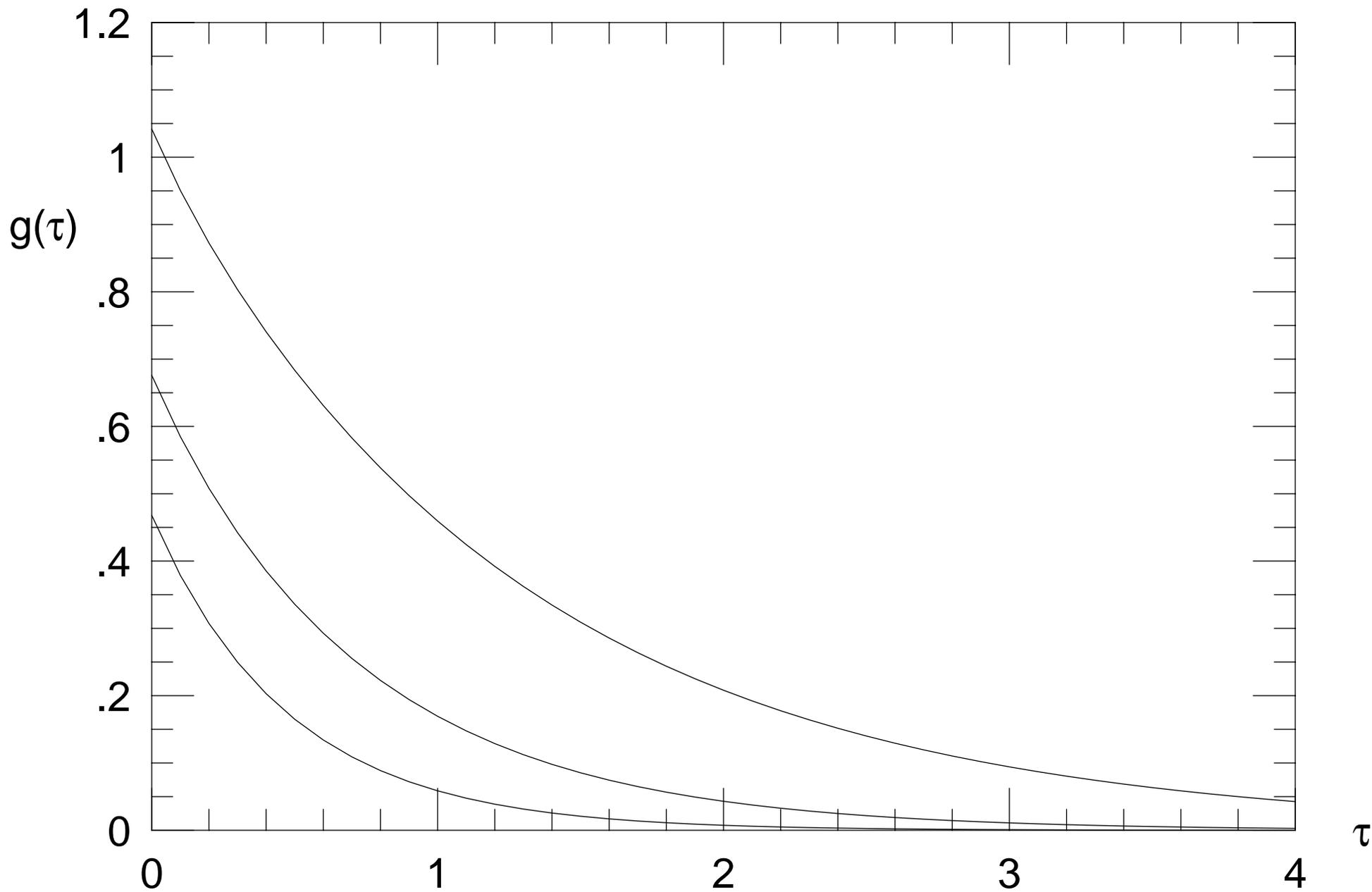

Figure 2

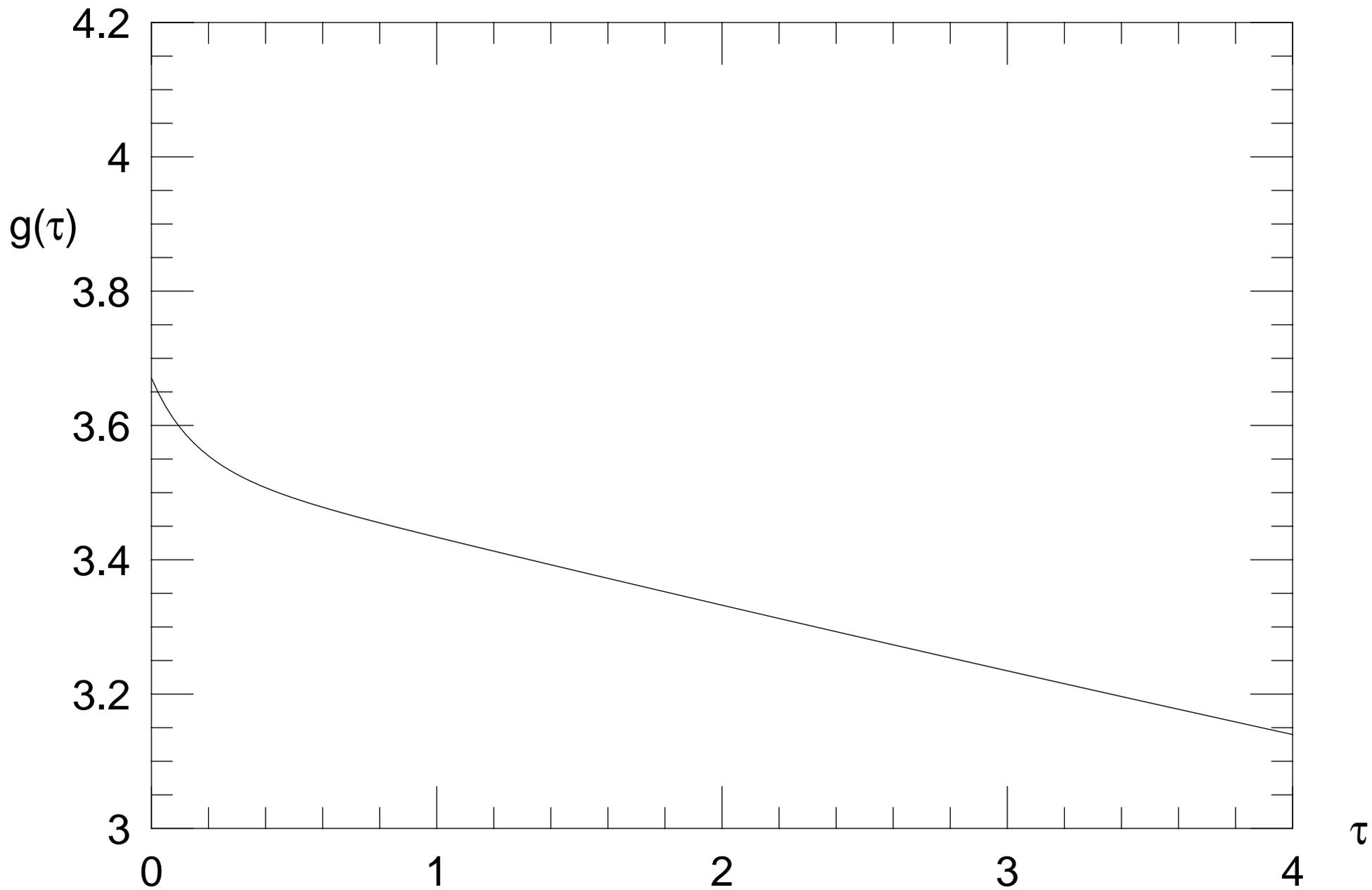

Figure 3